\documentclass[aps,prl,reprint,superscriptaddress]{revtex4-2}
\usepackage{amsfonts}
\usepackage{amssymb}
\usepackage{amsmath}
\usepackage{graphicx}
\usepackage{epsfig}
\usepackage{array}
\usepackage{braket}
\usepackage{multirow}
\usepackage[table,xcdraw]{xcolor}
\usepackage[colorlinks]{hyperref}
\usepackage{lineno}

\begin{document}

\title{Unconventional linear transverse exciton transport in valley-layer
coupling two-dimensional materials}
\author{Ci Li}
\email{lici@hnu.edu.cn}
\affiliation{School of Physics and Electronics, Hunan University, Changsha 410082, China}

\begin{abstract}
Valley-layer coupling (VLC) two-dimensional (2D) materials define a distinct
class of quantum systems in which valleys are related by crystal, rather
than time-reversal ($\mathcal{T}$) symmetry, enabling gate-controlled
valley-contrasted layer polarization. Here we extend this concept to
excitons. Using TiSiCo as a prototype VLC material, we show that a
perpendicular electric field $E_\perp$ controls exciton-dispersion
anisotropy and thereby generates unconventional linear transverse exciton
transport in both monolayer and twisted bilayer structures. In the
monolayer, this response is characterized by anisotropy-induced transverse
conductivities, arising from the antisymmetric combination of diagonal
elements in the conductivity tensor and strongly tunable by $E_\perp$. In
the twisted bilayer, symmetry additionally permits transverse responses from the symmetric part of the conductivity tensor. Since intralayer
excitons in different layers are connected by the F\"orster coupling, these
symmetric and antisymmetric responses coexist and compete with the recently
proposed $\mathcal{T}$-even layer Hall and Nernst exciton counterflow. This
interplay is highly tunable by twisted angle, temperature, and the direction
of the in-plane driving force, providing a route to disentangle distinct
transverse exciton transport signals experimentally. Our results establish $%
E_\perp$ as a powerful knob for controlling exciton transport via VLC and
identify VLC 2D materials as a promising platform for engineered excitonic
phenomena.
\end{abstract}

\maketitle




Excitons, the Coulomb-bound states of conduction electrons and valence
holes, are central to the optoelectronic properties of condensed matter
system \cite{Wang3,Wang1,Nov,Mak,Zhang,Gang}. In two-dimensional (2D)
materials, reduced dielectric screening yields large exciton binding
energies, typically on the order of hundreds of meV, ensuring stability even
at room temperature. Their small Bohr radii, usually a few nanometers,
further enhance light--matter coupling \cite{Wang3,Wang1,Nov,Zhang}. These
features make 2D semiconductors a leading platform for exciton physics,
particularly transition-metal dichalcogenides (TMDs), where valley-dependent
optical selection rules enable rich optical and transport phenomena \cite%
{Mak,Gang,Kai}. In twisted multilayer structures, the layer degree of
freedom provides an additional handle, leading to properties such as valley
pseudospin manipulation \cite{Yu,Lou}, electrically tunable exciton
anisotropy \cite{XWang,Yang}, interlayer excitons \cite{Xu,FWang,Yu1}, and
moir\'{e} excitons \cite{FWang,Yu1,Sey,Tra,Jin,Mak1,Jin1}.

The interplay between Berry curvature and quantum transport, which is
extensively studied in electronic systems \cite%
{Di,Niu,Nag,Xiao1,Xiao2,Mac1,Zhai,Xiao,JC}, has been extended to excitons 
\cite{Wang,Ong,Gol,She,Abr,Tong,Li2,Li3}. Both theory and experiment have
shown that neutral excitons can respond to external electric fields \cite%
{Gol,She} and exhibit Hall transport driven by excitonic Berry curvature 
\cite{Wang,Ong,Tong,Li2,Li3}, closely analogous to the anomalous Hall effect
of electrons \cite{Niu,Nag}. In parallel, nonlinear anomalous transport
recently predicted in twisted 2D electronic systems from symmetry
considerations \cite{Zhai,Xiao,JC} has also found excitonic counterparts in
bilayers \cite{Li2} with the F\"{o}rster coupling \cite{For,Seo,Kno,Li,Li1}.
These results highlight twisted 2D materials as a fertile setting for
uncovering transport phenomena rooted in chirality, layer structure, and
quantum geometry.

Most previous studies of transverse exciton transport in 2D materials have
focused on TMDs \cite{Wang,Ong,Gol,She,Tong,Kai1,Wu,Lun,Sav,Gao,Ros}, where
the response originates from exciton Berry curvature \cite%
{Wang,Ong,She,Tong,Kai,Wu,Sav,Gao,Ros} or from side-jump and skew-scattering
mechanisms \cite{Gol,Lun}. In these systems, however, opposite valleys are
related by time-reversal symmetry ($\mathcal{T}$), which prohibits purely
electrical generation of valley polarization by a gate field. Recently
proposed valley-layer-coupling (VLC) materials offer a fundamentally
different setting: distinct valleys are connected by crystal symmetry rather
than $\mathcal{T}$ \cite{Gao1,Ma,SAYang,Yao,Yong,Ji,Zhu}, enabling
gate-controlled valley-contrasted layer polarization in electronic states 
\cite{Ma,SAYang,Yong,Zhu}. This mechanism opens a new route for exciton
control, because it directly links an out-of-plane electric field to the
valley and layer structure underlying exciton dispersion \cite{SAYang,Ji}
and its anisotropy \cite{Tong,Li2,Li3}.

In this work, we show that VLC in 2D materials gives rise to qualitatively
new transverse exciton transport. Using TiSiCo (TSCO) as a prototype, we
demonstrate that a perpendicular electric field $E_{\perp }$ tunes
exciton-dispersion anisotropy and thereby induces unconventional linear
transverse exciton transport (LTET) in both monolayer and twisted bilayer
structures. In the monolayer, the response is characterized by
anisotropy-induced transverse conductivities, $\sigma _{\mathrm{tra}}^{%
\mathrm{AS}}$ and $\alpha _{\mathrm{tra}}^{\mathrm{AS}}$, originating from
the antisymmetric combination of diagonal elements in the conductivity
tensor and strongly tunable by $E_{\perp }$. In the twisted bilayer, F\"{o}%
rster coupling links intralayer excitons across the two layers, while
symmetry additionally permits transverse responses, $\sigma _{\mathrm{tra}}^{%
\mathrm{S}}$ and $\alpha _{\mathrm{tra}}^{\mathrm{S}}$, from the symmetric
part of the conductivity tensor. These responses coexist and compete with
the recently proposed $\mathcal{T}$-even layer Hall and Nernst exciton
counterflow \cite{Li2}. We further show that their interplay can be
efficiently controlled by twist angle, temperature, and the direction of the
in-plane driving force, providing a route to experimentally disentangle
distinct transverse exciton transport channels. Our results establish $%
E_{\perp }$ as a practical knob for engineering exciton transport in VLC
materials and identify this class of systems as a promising platform for
controllable excitonic phenomena.

\begin{figure}[tbp]
\begin{center}
\includegraphics[width=0.45\textwidth]{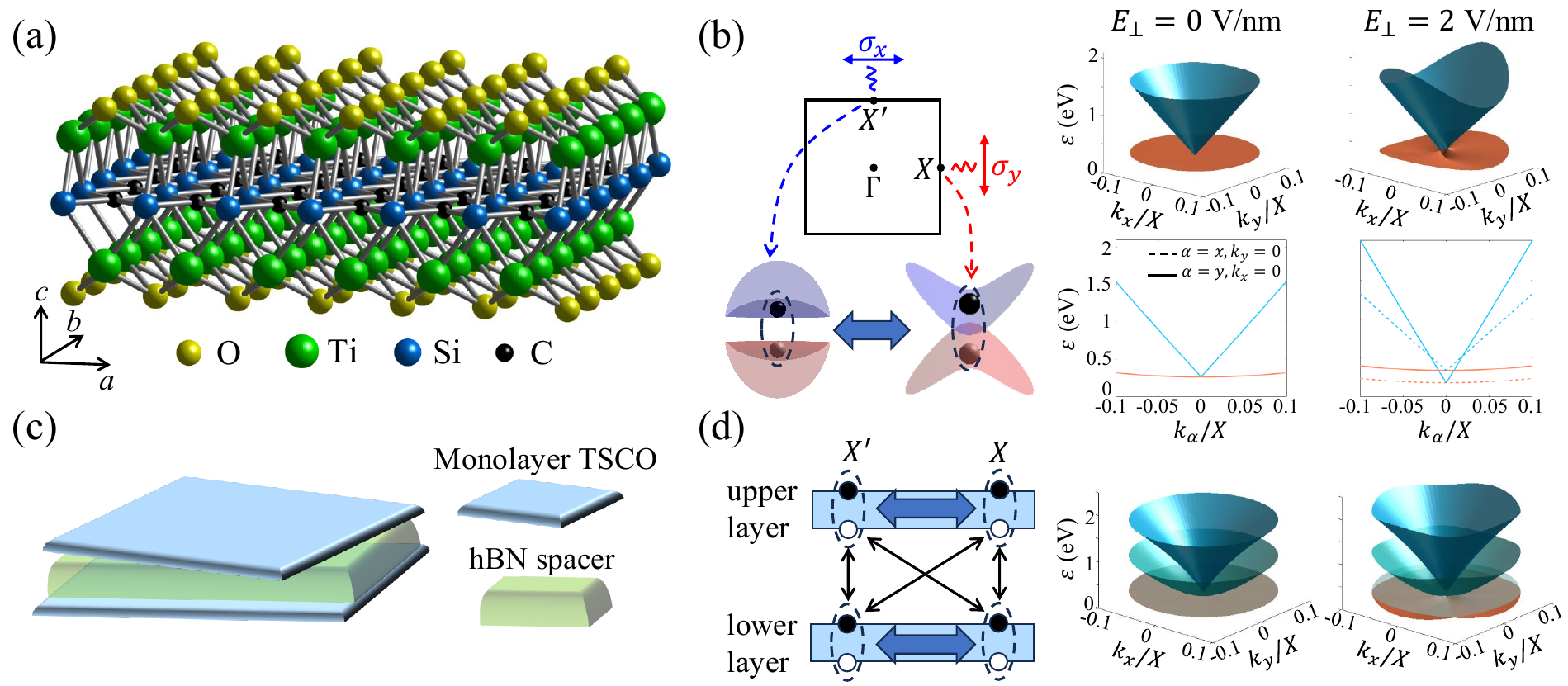}
\end{center}
\caption{(color online) (a) Crystal structure of monolayer TiSiCO (TSCO).
(b) Left: first Brillouin zone of monolayer TSCO and valley excitons at $X$
and $X^{\prime }$ with opposite optical selection rules; large blue double
arrows denote intervalley electron--hole exchange interaction. Right:
exciton dispersion of the monolayer effective Hamiltonian (\protect\ref{ml})
for different perpendicular electric fields $E_{\perp }$, shown in both 3D
view and selected momentum cuts. (c) Schematic of twisted bilayer TSCO
separated by a hexagonal boron nitride spacer. (d) Left: intralayer excitons
in twisted bilayer TSCO and their couplings; black double arrows denote F\"{o%
}rster coupling. Right: exciton dispersion of the twisted-bilayer effective
Hamiltonian (\protect\ref{tbl}) for different $E_{\perp }$ with twisted
angle $\protect\theta =\protect\pi /3$.}
\label{fig1}
\end{figure}

\section{Effective exciton Hamiltonians of monolayer and twisted bilayer TSCO%
}

Monolayer TSCO is a prototypical VLC 2D material \cite{Ma,SAYang,Yao,Ji}. It
belongs to the TiSiCO family of monolayer $X_{2}Y$CO$_{2}$ ($X=$ Ti, Zr, Hf; 
$Y=$ Si, Ge) \cite{SAYang,Yao,Dai,SAYang1} and was recently identified as
the first prototype cornertronic material \cite{Yao}. First-principles
calculations show that it is dynamically stable, thermally robust at high
temperature, and characterized by negligible spin-orbit coupling \cite%
{SAYang,Ji}. As illustrated in Fig. \ref{fig1}(a), the crystal consists of
five atomic layers with a thickness of about $4\,\mathrm{\mathring{A}}$,
corresponding to the separation between the two Ti layers, and has lattice
constants $a=b=2.817\,\mathrm{\mathring{A}}$\ \cite{SAYang,Yao,Ji}. The
monolayer preserves $\mathcal{T}$ and belongs to the $D_{2d}$ point group 
\cite{SAYang,Yao}, one of the 31 layer point groups \cite{Yao1}. Its
electric band structure reveals two valleys at the $X$ and $X^{\prime }$
points [Fig. \ref{fig1}(b)], which are related by crystal symmetries $S_{4z}$
and $C_{2,[110]}$ rather than by $\mathcal{T}$ \cite{SAYang,Yao}. Valley
excitons have likewise been predicted theoretically \cite{SAYang} and
confirmed numerically \cite{Ji}, with the two valleys supporting interlayer
excitons that exhibit opposite optical selection rules, as shown in Fig. \ref%
{fig1}(b).

Guided by the effective two-band valley model \cite{SAYang,Yao}, we
construct an effective exciton Hamiltonian in the valley-exciton basis with
the electron-hole (e-h) Coulomb exchange interaction as shown in Fig. \ref%
{fig1}(b) (see details in \cite{Supp}):%
\begin{eqnarray}
H_{ex}^{\mathrm{ML}}\left( \theta \right)  &=&\left[ 
\begin{array}{cc}
T_{X}\left( \theta \right)  & 0 \\ 
0 & T_{X^{\prime }}\left( \theta \right) 
\end{array}%
\right]   \label{ml} \\
&&+\left[ 
\begin{array}{cc}
J_{X,X}\left( \theta \right) +\varepsilon _{X} & J_{X,X^{\prime }}\left(
\theta \right)  \\ 
J_{X^{\prime },X}\left( \theta \right)  & J_{X^{\prime },X^{\prime }}\left(
\theta \right) +\varepsilon _{X^{\prime }}%
\end{array}%
\right] ,  \notag
\end{eqnarray}%
where $T_{\lambda =X,X^{\prime }}=\frac{\hbar ^{2}k_{x}^{2}}{2m_{ex,\lambda
}^{x}\left( \theta \right) }+\frac{\hbar ^{2}k_{y}^{2}}{2m_{ex,\lambda
}^{y}\left( \theta \right) }$ and $\hbar \boldsymbol{k}=\hbar \left(
k_{x},k_{y}\right) =\hbar k\left( \cos \varphi ,\sin \varphi \right) $ is
the exciton centre-of-mass (COM) momentum. $m_{ex,\lambda }^{\gamma
=x,y}\left( \theta \right) =m_{\gamma ,\lambda }^{e}\left( \theta \right)
+m_{\gamma ,\lambda }^{h}\left( \theta \right) $ is the exciton mass with
effective electron (hole) mass $m_{\gamma ,\lambda }^{e\left( h\right)
}\left( \theta \right) $ in valley $\lambda $. The exchange terms are
approximately 
\begin{eqnarray*}
J_{X,X}\left( \theta \right)  &\approx &J_{X}\left( \theta \right) \frac{%
k\sin ^{2}\left( \varphi -\theta \right) }{\left\vert X\right\vert }, \\
J_{X,X}\left( \theta \right)  &\approx &J_{X^{\prime }}\left( \theta \right) 
\frac{k\cos ^{2}\left( \varphi -\theta \right) }{\left\vert X\right\vert },
\\
J_{X,X^{\prime }}\left( \theta \right)  &=&J_{X^{\prime },X}\left( \theta
\right)  \\
&\approx &\sqrt{J_{X}\left( \theta \right) J_{X^{\prime }}\left( \theta
\right) }\frac{k\cos \left( \varphi -\theta \right) \sin \left( \varphi
-\theta \right) }{\left\vert X\right\vert }/\left\vert X\right\vert ,
\end{eqnarray*}%
where $\left\vert X\right\vert =\pi /a$ is the half width of the first
Brillouin zone (BZ), and $J_{\lambda }\left( \theta \right) $ is the energy
term set by the e-h exchange \cite{Supp}. The valley exciton energies at $k=0
$ are $\varepsilon _{X}\approx 2\left( \Delta +\alpha _{E}E_{\perp }\right) $%
, $\varepsilon _{X^{\prime }}=2\left( \Delta -\alpha _{E}E_{\perp }\right) $
with $\Delta \approx 0.133\,\mathrm{eV}$ and $\alpha _{E}\approx -0.2\,%
\mathrm{\mathring{A}}$ from DFT results \cite{SAYang}; the exciton binding
energy is omitted for simplicity. The angle $\theta $ donates the relative
angle between the crystal orientation $\boldsymbol{a}$ with the $x$ axis of
chosen Cartesian coordinate, and can be set to zero for a monolayer. The
exciton dispersion with different values of $E_{\perp }$ has been shown in
Fig. \ref{fig1}(b). As shown in the same plot, a finite $E_{\perp }$
strongly enhances the anisotropy of the exciton dispersion, while the $%
E_{\perp }=0$ case remains nearly isotropic. In details, both the magnitude
and sign of $E_{\perp }$ can provide efficient control of the exciton
spectrum (see Fig. S3 in \cite{Supp}).

For twisted bilayer TSCO [Fig. \ref{fig1}(c)], the exciton Hamiltonian [Fig. %
\ref{fig1}(d)] consists of two monolayer blocks coupled by interlayer F\"{o}%
rster transfer between intralayer excitons in different layers \cite{Note2}%
\begin{equation}
H_{ex}^{\mathrm{tBL}}=\left( 
\begin{array}{cc}
H_{ex}^{\mathrm{ML},1} & 0 \\ 
0 & H_{ex}^{\mathrm{ML},2}%
\end{array}%
\right) +\left( 
\begin{array}{cc}
0 & H_{\mathrm{inter}}^{1,2} \\ 
H_{\mathrm{inter}}^{2,1} & 0%
\end{array}%
\right) ,  \label{tbl}
\end{equation}%
We take layer $1$ as untwisted (bottom layer) and layer $2$ rotated by
twisted angle $\theta $ [Fig. \ref{fig1}(c)], so that $H_{ex}^{\mathrm{ML}%
,1}\equiv H_{ex}^{\mathrm{ML}}\left( 0\right) $ and $H_{ex}^{\mathrm{ML}%
,2}\equiv H_{ex}^{\mathrm{ML}}\left( \theta \right) $. The interlayer
coupling is%
\begin{equation*}
H_{\mathrm{inter}}^{1,2}=\left( 
\begin{array}{cc}
J_{X,X}^{1,2} & J_{X,X^{\prime }}^{1,2} \\ 
J_{X^{\prime },X}^{1,2} & J_{X^{\prime },X^{\prime }}^{1,2}%
\end{array}%
\right) =\left( H_{\mathrm{inter}}^{2,1}\right) ^{\dagger },
\end{equation*}%
with \cite{Supp} 
\begin{eqnarray*}
J_{X,X}^{1,2} &\approx &\sqrt{J_{X}\left( 0\right) J_{X}\left( \theta
\right) }\frac{ke^{-kz}\sin \varphi \sin \left( \varphi -\theta \right) }{%
\left\vert X\right\vert }, \\
J_{X^{\prime },X^{\prime }}^{1,2} &\approx &\sqrt{J_{X^{\prime }}\left(
0\right) J_{X^{\prime }}\left( \theta \right) }\frac{ke^{-kz}\cos \varphi
\cos \left( \varphi -\theta \right) }{\left\vert X\right\vert }, \\
J_{X,X^{\prime }}^{1,2} &\approx &\sqrt{J_{X}\left( 0\right) J_{X^{\prime
}}\left( \theta \right) }\frac{ke^{-kz}\sin \varphi \cos \left( \varphi
-\theta \right) }{\left\vert X\right\vert }, \\
J_{X^{\prime },X}^{1,2} &\approx &\sqrt{J_{X^{\prime }}\left( 0\right)
J_{X}\left( \theta \right) }\frac{ke^{-kz}\cos \varphi \sin \left( \varphi
-\theta \right) }{\left\vert X\right\vert },
\end{eqnarray*}%
and $z$ means the interlayer space that is normally around $1\,\mathrm{nm}$.
As shown in Fig. \ref{fig1}(d), the twisted bilayer also exhibits
electrically tunable exciton-dispersion anisotropy. In contrast to the
monolayer, however, the anisotropy is now controlled jointly by $E_{\perp }$
and $\theta $ (see Fig. S4 in \cite{Supp}), demonstrating that the layer
degree of freedom provides an additional and highly flexible handle for
engineering excitonic properties in VLC systems.

\begin{figure}[tbp]
\begin{center}
\includegraphics[width=0.45\textwidth]{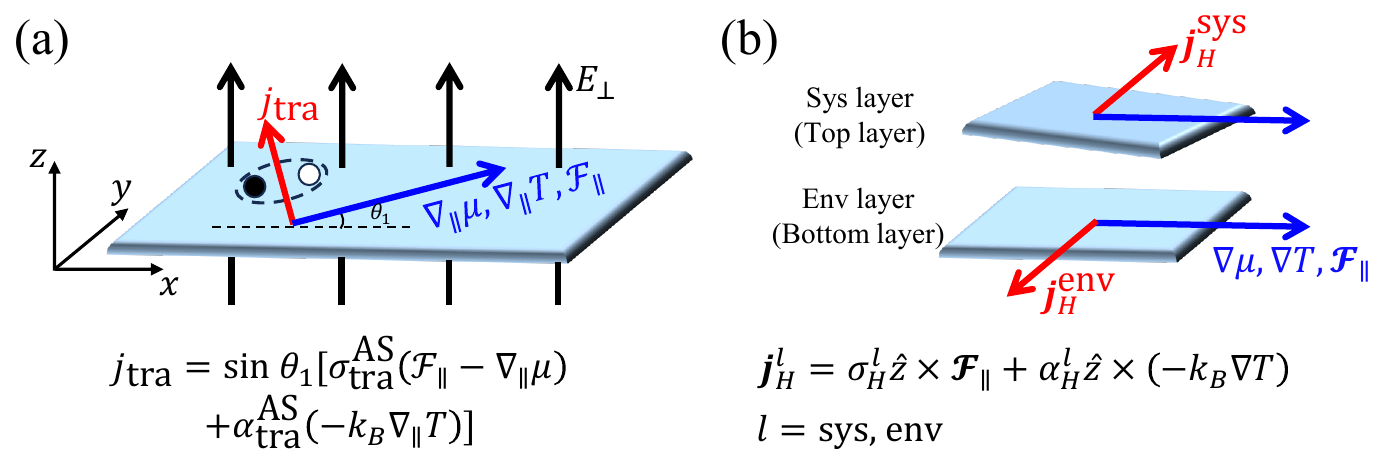}
\end{center}
\caption{(color online) (a) Schematic of linear transverse exciton transport
in a monolayer. (b) Schematic of the time-reversal even layer Hall (Nernst)
exciton counterflow in a twisted bilayer.}
\label{fig2}
\end{figure}

\begin{figure*}[tbp]
\begin{center}
\includegraphics[width=0.8\textwidth]{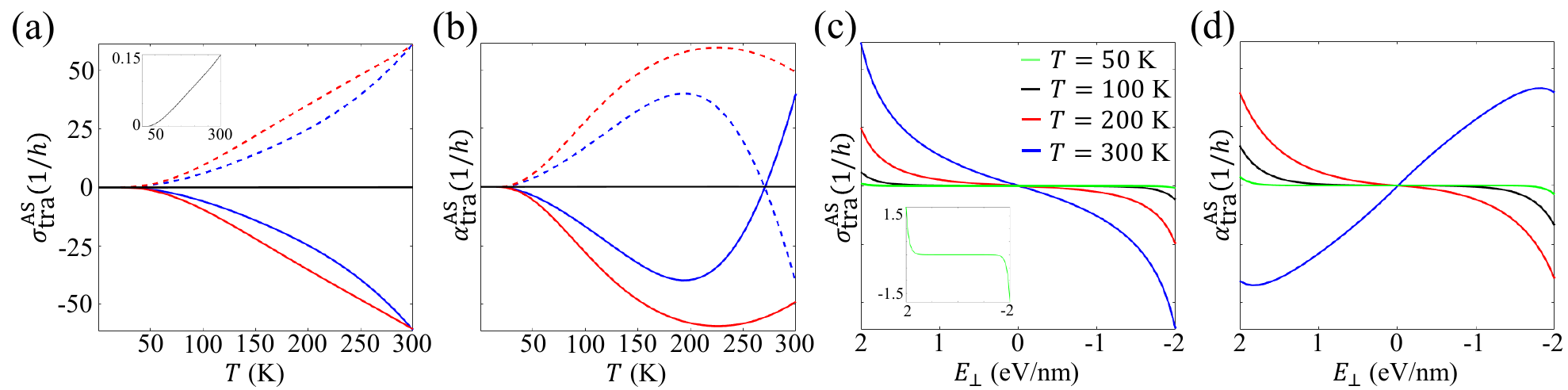}
\end{center}
\caption{(color online) (a)-(d) Numerical calculation of $\protect\sigma _{%
\mathrm{tra}}^{\mathrm{AS}}$ and $\protect\alpha _{\mathrm{tra}}^{\mathrm{AS}%
}$ [Eq. (\protect\ref{TC})] for monolayer exciton Hamiltonian (\protect\ref%
{ml}), as functions of temperature $T$ and perpendicular electric field $%
E_{\perp }$. Insets in (a) and (c) show enlarged views near $E_{\perp }=0$
and at $T=50\,K$, respectively. In (a) and (b), blue solid (dashed) curves
correspond to $E_{\perp }=-2(2)\,\mathrm{eV/nm}$, red solid (dashed) curves
to $E_{\perp }=-1(1)\,\mathrm{eV/nm}$, and black curves to $E_{\perp }=0$%
\thinspace $\mathrm{eV/nm}$. $\min (\protect\varepsilon _{n})-\protect\mu %
=10\,\mathrm{meV}$; In (c) and (d), $\protect\mu =0.176\,\mathrm{eV}$. $%
\protect\tau =1\,\mathrm{ps}$ for all plots.}
\label{fig3}
\end{figure*}

\section{LTET in monolayer TSCO}

The electric-field tunability of exciton-dispersion anisotropy in monolayer
TSCO enables unconventional LTET \cite{Li3} driven by an in-plane force on
excitons \cite{Note4}, as shown in Fig. \ref{fig2}(a). Because the $D_{2d}$
symmetry of TSCO enforces vanishing exciton Berry curvature \cite{Li3}. The
current of LTET, $j_{\mathrm{tra}}$, arises entirely from crystalline
anisotropy. We therefore characterize it by the anisotropy-induced
transverse conductivities [Fig. \ref{fig2}(a)],%
\begin{equation}
\sigma _{\mathrm{tra}}^{\mathrm{AS}}\equiv \frac{\sigma _{yy}^{1}-\sigma
_{xx}^{1}}{2},\alpha _{\mathrm{tra}}^{\mathrm{AS}}\equiv \frac{\alpha
_{yy}^{1}-\alpha _{xx}^{1}}{2},  \label{TC}
\end{equation}%
arising from the antisymmetric combination of diagonal elements in the
conductivity tensor \cite{Supp}, where%
\begin{eqnarray*}
\sigma _{\gamma \gamma }^{1} &=&\frac{\tau }{k_{B}T}\sum_{n}\int \frac{d^{2}%
\boldsymbol{k}}{\left( 2\pi \right) ^{2}}v_{\gamma ,n}^{2} \\
&&\times \left\{ f_{n}^{0}\left( \boldsymbol{k}\right) +\left[
f_{n}^{0}\left( \boldsymbol{k}\right) \right] ^{2}\right\} ,
\end{eqnarray*}%
\begin{eqnarray*}
\alpha _{\gamma \gamma }^{1} &=&\frac{\tau }{k_{B}T}\sum_{n}\int \frac{d^{2}%
\boldsymbol{k}}{\left( 2\pi \right) ^{2}}v_{\gamma ,n}^{2} \\
&&\times \frac{\varepsilon _{n}-\mu }{k_{B}T}\left\{ f_{n}^{0}\left( 
\boldsymbol{k}\right) +\left[ f_{n}^{0}\left( \boldsymbol{k}\right) \right]
^{2}\right\} .
\end{eqnarray*}%
Here $f_{n}^{0}\left( \boldsymbol{k}\right) =1/\left[ e^{\left( \varepsilon
_{n}-\mu \right) /k_{B}T}-1\right] $ is the equilibrium Bose-Einstein
distribution function. $v_{\gamma ,n}\equiv \frac{\partial \varepsilon
_{n}\left( \boldsymbol{k}\right) }{\hbar \partial k_{\beta }}$ means the
band velocity of the excitonic band $n$. $\tau $ is the relaxation time in
the first-order relaxation-time approximation \cite{Li3}. Unlike
conventional Hall and Nernst responses in electronic systems, which are $%
\mathcal{T}$-odd and independent to the driving-field direction \cite%
{Niu,Ong,Ons}, $\sigma _{\mathrm{tra}}^{\mathrm{AS}}$ and $\alpha _{\mathrm{%
tra}}^{\mathrm{AS}}$ are $\mathcal{T}$-even and originate from crystal
anisotropy. Recently, similar $\mathcal{T}$-even transverse responses have
been discussed in analogy to the spin Hall effect in antiferromagnets \cite%
{Jun,Ohn,Liu}.

The numerical results are shown in Fig. \ref{fig3}. The dependence on the
sign of $E_{\perp }$ can be qualitatively understood from the valley exciton
energies $\varepsilon _{X}$ and $\varepsilon _{X^{\prime }}$ at $k=0$. At $%
E_{\perp }=0$, the two valleys are crystallographically inequivalent but
nearly degenerate in energy ($\varepsilon _{X}\approx \varepsilon
_{X^{\prime }}$). As a result, the anisotropy of the exciton Hamiltonian
[Eq. (\ref{ml})] is reduced relative to that of the underlying crystal,
yielding finite but much smaller transverse responses than at nonzero field.
For $E_{\perp }\neq 0$, the valley degeneracy is lifted, i.e., $\varepsilon
_{X}\neq \varepsilon _{X^{\prime }}$, which enhances the anisotropy and
produces sizable conductivities. Reversing the field exchanges $\varepsilon
_{X}$ and $\varepsilon _{X^{\prime }}$, leading to $\sigma _{yy}^{1}$ $%
\left( \alpha _{yy}^{1}\right) $ $\leftrightarrow \sigma _{xx}^{1}$ $\left(
\alpha _{xx}^{1}\right) $ (see Fig. S5 in \cite{Supp}), and hence a reversal
of the LTET. The Hall-like conductivity $\sigma _{\mathrm{tra}}^{\mathrm{AS}%
} $ varies nearly monotonically with temperature $T$ [Fig. \ref{fig3}(a)],
whereas the Nernst-like conductivity $\alpha _{\mathrm{tra}}^{\mathrm{AS}}$
exhibits pronounced nonmonotonic behavior [Fig. \ref{fig3}(b)]. This
contrast also appears in their electric-field dependence [Figs. \ref{fig3}%
(c) and (d)]. Most importantly, both responses remain highly sensitive to $%
E_{\perp }$ over a broad temperature range, identifying monolayer TSCO as a
promising platform for electrically controlled LTET.

\begin{figure*}[tbp]
\begin{center}
\includegraphics[width=0.9\textwidth]{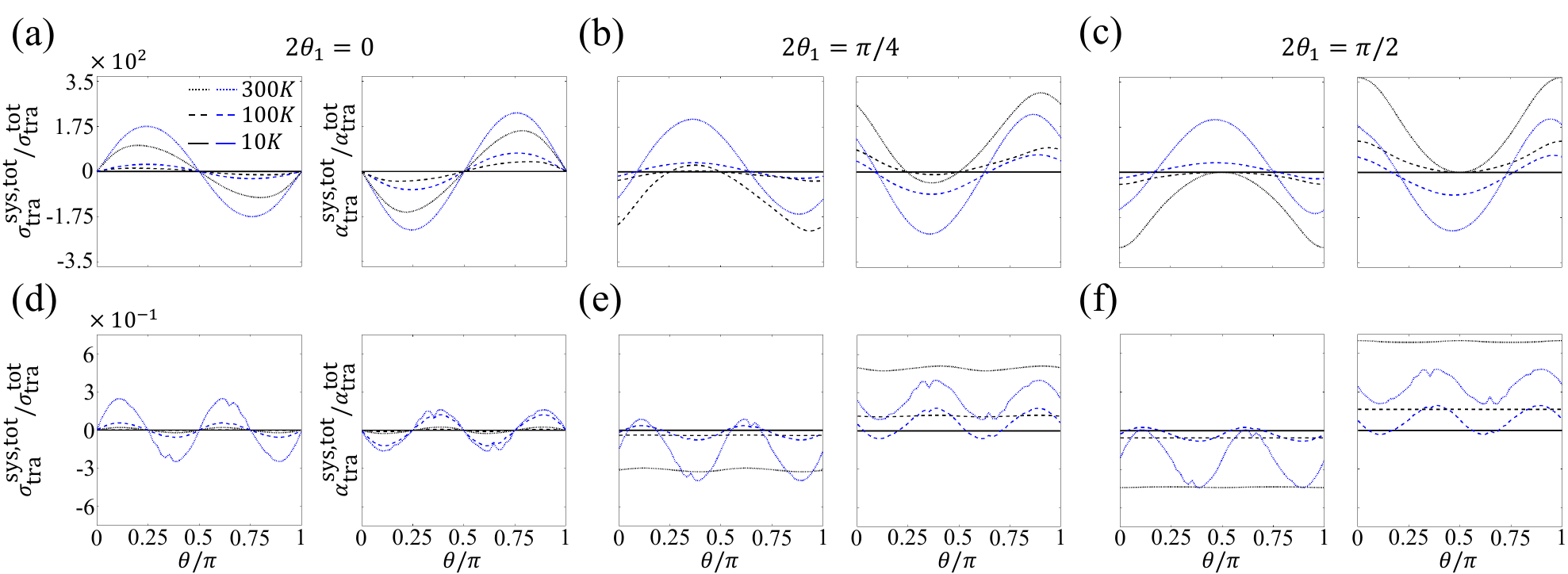}
\end{center}
\caption{(color online) (a) Left: Numerical calculations of total transverse
conductivity in system layer for twisted bilayer exciton Hamiltonian (%
\protect\ref{tbl}), $\protect\sigma _{\mathrm{tra}}^{\mathrm{sys},\mathrm{tot%
}}\equiv \protect\sigma _{H}^{\mathrm{sys}}+\protect\sigma _{\mathrm{tra}}^{%
\mathrm{sys},\mathrm{AS}}\sin (2\protect\theta _{1})+\protect\sigma _{%
\mathrm{tra}}^{\mathrm{sys},\mathrm{S}}\cos (2\protect\theta _{1})$ [Eqs. (%
\protect\ref{TC}) and (\protect\ref{TRE})] (blue curves), and $\protect%
\sigma _{\mathrm{tra}}^{\mathrm{tot}}=\protect\sigma _{\mathrm{tra}}^{%
\mathrm{sys},\mathrm{tot}}+\protect\sigma _{\mathrm{tra}}^{\mathrm{env},%
\mathrm{tot}}$ (black curves), as functions of twisted angle $\protect\theta 
$ and temperature $T$, in the unit of $1/h$. Right: Corresponding results
for $\protect\alpha _{\mathrm{tra}}^{\mathrm{sys},\mathrm{tot}}$ and $%
\protect\alpha _{\mathrm{tra}}^{\mathrm{tot}}$. Here perpendicular electric
field $E_{\perp }=-2\,\mathrm{eV/nm}$. (b) and (c): Similar plots as in (a),
but for different in-plane driving force direction $\protect\theta _{1}$.
(d)-(f): Similar plots as (a)-(c), but for $E_{\perp }=0\,\mathrm{eV/nm}$.
In all plots, $\min (\protect\varepsilon _{n})-\protect\mu =10\,\mathrm{meV}$
and $\protect\tau =1\,\mathrm{ps}$.}
\label{fig4}
\end{figure*}

\begin{figure}[tbp]
\begin{center}
\includegraphics[width=0.45\textwidth]{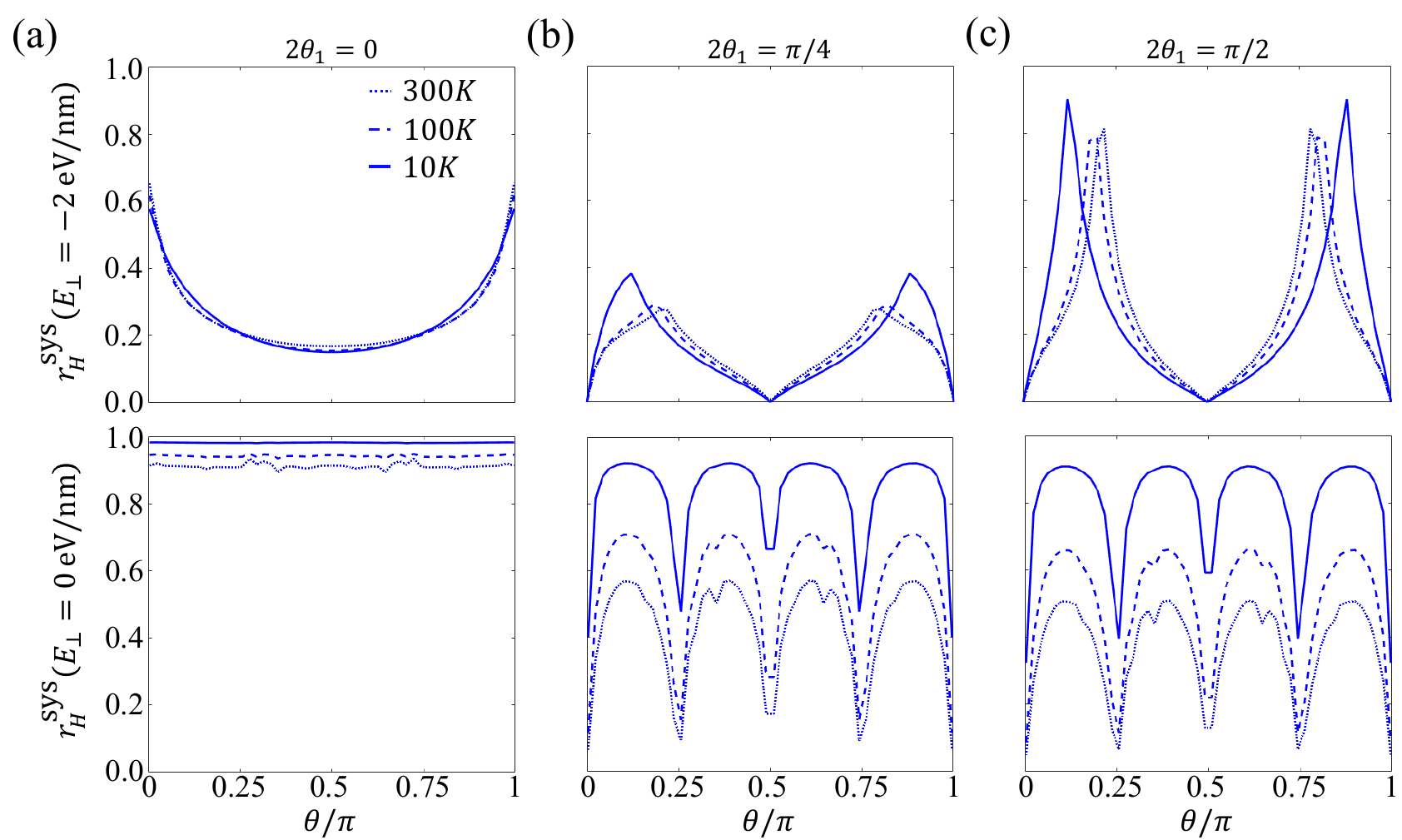}
\end{center}
\caption{(color online) (a) Numerical calculation of Hall fractions $r_{H}^{%
\mathrm{sys}}\equiv |\protect\sigma _{H}^{\mathrm{sys}}|/(|\protect\sigma %
_{H}^{\mathrm{sys}}|+|\protect\sigma _{\mathrm{tra}}^{\mathrm{sys},\mathrm{AS%
}}\sin(2\protect\theta_1)|+|\protect\sigma _{\mathrm{tra}}^{\mathrm{sys},%
\mathrm{S}}\cos(2\protect\theta_1)|)$ for different perpendicular electric
field $E_{\perp }$. (b) and (c): Similar plots as in (a), but for different
in-plane driving force direction $\protect\theta_1$. $\min (\protect%
\varepsilon _{n})-\protect\mu =10\,\mathrm{meV}$ and $\protect\tau =1\,%
\mathrm{ps}$ in all plots.}
\label{fig5}
\end{figure}

\section{LTET in twisted bilayer TSCO}

The layer degree of freedom and chirality of twisted bilayer TSCO, which
lacks any mirror symmetry, allow a $\mathcal{T}$-even layer Hall (Nernst)
exciton counterflow (TREHCF) \cite{Zhai,Xiao,Li2}, as shown in Fig. \ref%
{fig2}(b). Within semiclassical transport theory \cite{Li2}, the exciton
current in the top (system) and bottom (environment) layers is 
\begin{equation*}
\boldsymbol{j}^{\mathrm{sys/env}}=\sum_{n}\int \frac{d^{2}\boldsymbol{k}}{%
\left( 2\pi \right) ^{2}}f_{n}\left( \boldsymbol{k}\right) \boldsymbol{v}%
_{n}^{\mathrm{sys/env}}\left( \boldsymbol{k}\right) ,
\end{equation*}
where $f_{n}\left( \boldsymbol{k}\right) \approx f_{n}^{0}\left( \boldsymbol{%
k}\right) -\tau \frac{df_{n}^{0}\left( \boldsymbol{k}\right) }{dt}$ is the
non-equilibrium distribution of band $n$ in the relaxation-time
approximation. $\boldsymbol{v}_{n}^{\mathrm{sys/env}}\left( \boldsymbol{k}%
\right) $ is the layer-projected band velocity \cite{Zhai,Supp}. Onsager
relation enforces $\boldsymbol{j}^{\mathrm{sys}}=-\boldsymbol{j}^{\mathrm{env%
}}$ \cite{Zhai,Li2}. The corresponding Hall and Nernst conductivities for
TREHCF of the system layer are $\sigma _{H}^{\mathrm{sys}}=\tau \mathcal{V}%
/\hbar $ and $\alpha _{H}^{\mathrm{sys}}=\tau \mathcal{V}_{\mathrm{Ner}%
}/\hbar $, with%
\begin{eqnarray}
\mathcal{V} &=&\sum_{n}\int \frac{d^{2}\boldsymbol{k}}{\left( 2\pi \right)
^{2}}f_{n}^{0}\left( \boldsymbol{k}\right) \omega _{n}\left( \boldsymbol{k}%
\right) ,  \label{TRE} \\
\mathcal{V}_{\mathrm{Ner}} &=&\sum_{n}\int \frac{d^{2}\boldsymbol{k}}{\left(
2\pi \right) ^{2}}\frac{\varepsilon _{n}-\mu }{k_{B}T}f_{n}^{0}\left( 
\boldsymbol{k}\right) \omega _{n}\left( \boldsymbol{k}\right) ,  \notag
\end{eqnarray}%
where $\omega _{n}\left( \boldsymbol{k}\right) =\frac{1}{2}\left[ \nabla _{%
\boldsymbol{k}}\times \boldsymbol{v}_{n}^{\mathrm{sys}}\left( \boldsymbol{k}%
\right) \right] $ is the $\boldsymbol{k}$-space vorticity of the layer
current \cite{Zhai}, reflecting the quantum geometry of layer-hybridized
exciton states in the chiral bilayer. Although TREHCF produces no net
current for the full bilayer, it yields opposite Hall currents in the two
layers and can therefore be detected through layer-resolved probes \cite%
{Zhai,Li2}.

The crystalline-anisotropy-driven LTET identified in the monolayer persists
in the twisted bilayer and enters at the same order in the relaxation-time
expansion. For each layer, the transverse response therefore contains the $%
\mathcal{T}$-even layer-counterflow contribution of Eq. (\ref{TRE}) and the
anisotropy-driven contribution of Eq. (\ref{TC}). Owing to $D_{2d}$ symmetry
of TSCO, the twisted bilayer also allows additional transverse responses, $%
\sigma _{\mathrm{tra}}^{l,\mathrm{S}}$ and $\alpha _{\mathrm{tra}}^{l,%
\mathrm{S}}$ ($l={\mathrm{sys,env}}$), from the symmetric part of the
conductivity tensor, i.e., the symmetric combination of non-diagonal
elements (see \cite{Li3} and \cite{Supp} for details). The layer-resolved
total transverse conductivity $\sigma _{\mathrm{tra}}^{l,\mathrm{tot}}$ ($%
\alpha _{\mathrm{tra}}^{l,\mathrm{tot}}$) thus contains all three
contributions, and the bilayer response is $\sigma _{\mathrm{tra}}^{\mathrm{%
tot}}=\sigma _{\mathrm{tra}}^{\mathrm{sys},\mathrm{tot}}+\sigma _{\mathrm{tra%
}}^{\mathrm{env},\mathrm{tot}}$, with an analogous definition for $\alpha _{%
\mathrm{tra}}^{\mathrm{tot}}$. As shown in Fig. \ref{fig4}, Both $\sigma _{%
\mathrm{tra}}^{\mathrm{sys},\mathrm{tot}}$ and $\alpha _{\mathrm{tra}}^{%
\mathrm{sys},\mathrm{tot}}$, as well as the full bilayer responses $\sigma _{%
\mathrm{tra}}^{\mathrm{tot}}$ and $\alpha _{\mathrm{tra}}^{\mathrm{tot}}$,
are periodic in the twisted angle $\theta $, with the periodicity controlled
by $E_{\perp }$. At $E_{\perp }=0$, the two valleys are exciton-energy
degenerate, giving rise to an effective $\pi /2$ periodicity despite the
underlying $D_{2d}$ crystal symmetry. A finite $E_{\perp }$ lifts this
degeneracy and restores the symmetry-allowed $\pi $ periodicity, consistent
with the monolayer case. Figure \ref{fig4} further shows a net $\theta $%
-dependent LTET current for the entire bilayer. This net response is
dominated by the symmetric and antisymmetric anisotropic terms, because the
total Hall contribution vanishes exactly, $\sigma _{H}=\sigma _{H}^{\mathrm{%
sys}}+\sigma _{H}^{\mathrm{env}}=0$, as required by the Onsager relation 
\cite{Nag,Zhai,Li3,Supp}. Increasing $|E_{\perp }|$ generally enhances all
transverse conductivities and sharpens their angular dependence, where the
sign of $E_{\perp }$ can control the direction of these responses (see Fig.
S6 in \cite{Supp}). Except near special twisted angles, these responses
increase monotonically with temperature and remain sizable---of order
several $1/h$---around 100$\,K$ at finite $E_{\perp }$, making them
promising for experimental detection.


Figure \ref{fig5} shows that the observable layer-resolved response is
determined by the competition among three contributions within each layer:
the layer-counterflow term, the anisotropic term from the antisymmetric part of diagonal elements in the conductivity tensor, and $\sigma_{\mathrm{tra}}^{l,\mathrm{S}}$ from
the symmetric part of non-diagonal elements. This competition is conveniently quantified by the Hall
fraction, 
\begin{equation}
r_{H}^{l}\equiv\frac{|\sigma _{H}^{\mathrm{sys}}|}{|\sigma _{H}^{\mathrm{sys}%
}|+|\sigma _{\mathrm{tra}}^{\mathrm{sys},\mathrm{AS}}\sin(2\theta_1)|+|%
\sigma _{\mathrm{tra}}^{\mathrm{sys},\mathrm{S}}\cos(2\theta_1)|}.
\end{equation}
In particular, $E_{\perp }$ provides an efficient means to tune their
relative weight as functions of $\theta $ and $\theta_1$, while increasing
temperature generally reduces $r_{H}^{l}$, suppressing the contribution from
TREHCF. The thermoelectric counterparts exhibit the same qualitative
behavior (see Fig. S7 in \cite{Supp}). These results establish twisted
bilayer TSCO as a versatile platform in which distinct $\mathcal{T}$-even
transverse exciton transport channels can be disentangled by electric field,
twisted angle, direction of in-plane driving force, and temperature.

\begin{figure}[tbp]
\begin{center}
\includegraphics[width=0.45\textwidth]{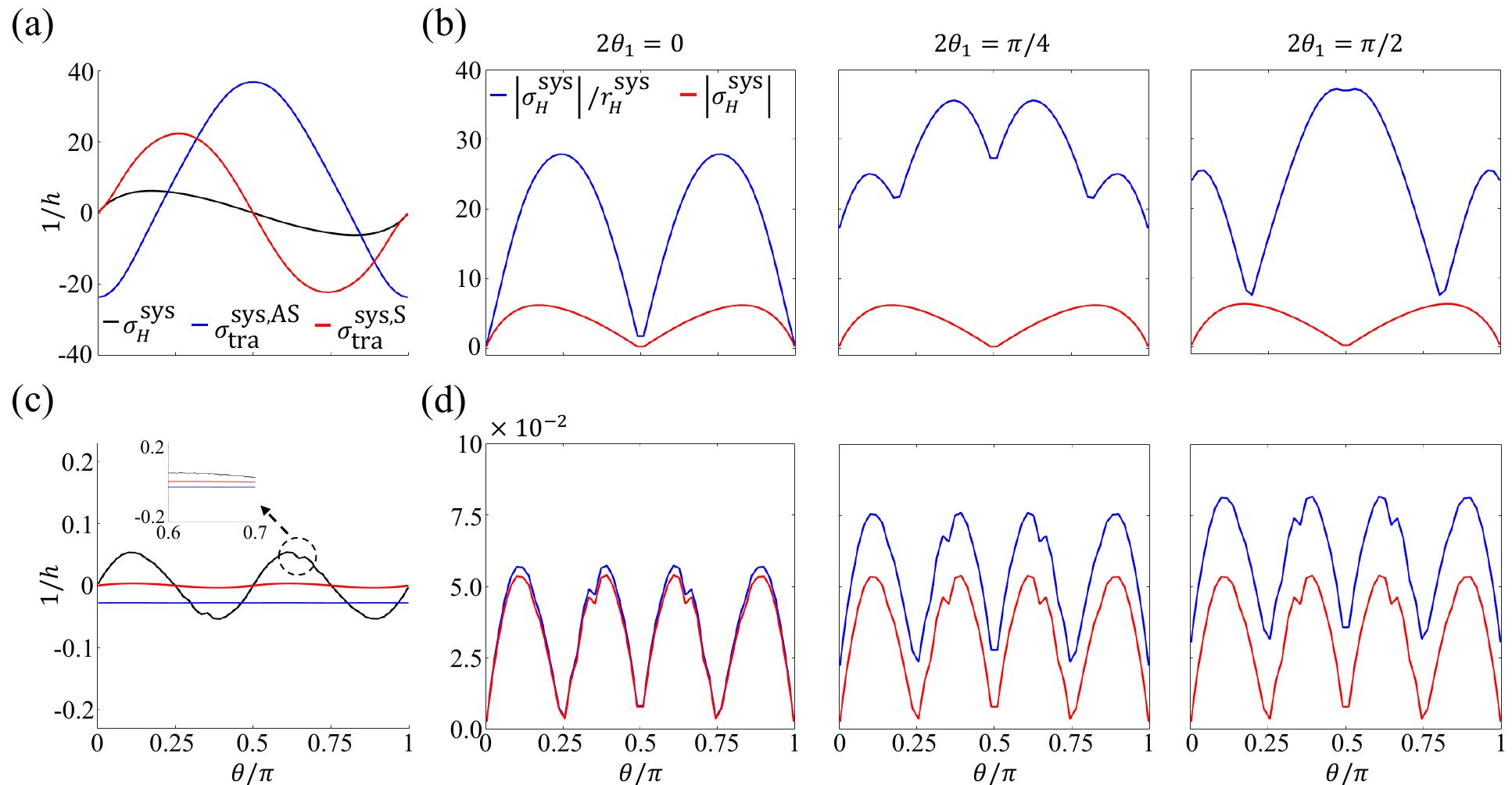}
\end{center}
\caption{(color online) (a) Numerical calculation of layer-resolved
transverse exciton conductivities $\protect\sigma _{H}^{\mathrm{sys}}$, $%
\protect\sigma _{\mathrm{tra}}^{\mathrm{sys},\mathrm{AS}}$, and $\protect%
\sigma _{\mathrm{tra}}^{\mathrm{sys},\mathrm{S}}$ for perpendicular electric
field $E_{\perp }=-2\,\mathrm{eV/nm}$ and temperature $T=100\,K$. (b)
Corresponding quantities $|\protect\sigma _{H}^{\mathrm{sys}}|/r_{H}^{%
\mathrm{sys}}=|\protect\sigma _{H}^{\mathrm{sys}}|+|\protect\sigma _{\mathrm{%
tra}}^{\mathrm{sys},\mathrm{AS}}\sin (2\protect\theta _{1})|+|\protect\sigma %
_{\mathrm{tra}}^{\mathrm{sys},\mathrm{S}}\cos (2\protect\theta _{1})|$ and $|%
\protect\sigma _{H}^{\mathrm{sys}}|$, for different in-plane driving force
directions $\protect\theta _{1}$. (c) and (d): Similar plots as (a) and (b),
but for $E_{\perp }=0\,\mathrm{eV/nm}$. $\min (\protect\varepsilon _{n})-%
\protect\mu =10\,\mathrm{meV}$ and $\protect\tau =1\,\mathrm{ps}$ in all
plots.}
\label{fig6}
\end{figure}

The apparent discontinuities with $\theta$ in $\sigma^{\mathrm{sys},\mathrm{%
tot}}_{\mathrm{tra}}$ and $\alpha^{\mathrm{sys},\mathrm{tot}}_{\mathrm{tra}}$
in Figs. \ref{fig4}(d)-(f), and in $r^{\mathrm{sys}}_{H}$ in Fig. \ref{fig5}%
, can be understood from the layer-resolved transverse responses shown in
Fig. \ref{fig6}. For $E_{\perp }\neq 0$, the underlying transverse responses
vary continuously with $\theta$; the discontinuities arise only after taking
absolute values, which renders both $|\sigma _{H}^{\mathrm{sys}}|/r_{H}^{%
\mathrm{sys}}$ and $|\sigma _{H}^{\mathrm{sys}}|$ discontinuous at specific
angles [Figs. \ref{fig6}(a) and (b)]. At $E_{\perp }=0$, the effect is
instead tied to the $\theta$-dependence of $\sigma_{H}^{\mathrm{sys}}$
[Figs. \ref{fig6}(c) and (d)]: at finite temperature, $\sigma_{H}^{\mathrm{%
sys}}$ exhibits weak discontinuities---more precisely, narrow oscillatory
features over a small range of $\theta$, as highlighted in the inset in Fig. %
\ref{fig6}(c)---which produce the behavior seen in Figs. \ref{fig4} and \ref%
{fig5}.

\section{Outlook and discussions}

Recent work has shown that at sufficiently high temperature or in the
presence of strong disorder, impurity- and phonon-induced skew-scattering
and, in particular, side-jump processes can substantially modify the exciton
valley Hall effect; notably, the side-jump contribution may partially
compensate the Berry-curvature-induced anomalous velocity \cite{Gol,Gol1,Dya}%
. In that regime, the valley Hall current remains linear in the in-plane
driving force, but the conductivity can become dominated by skew scattering
and scale approximately linearly with temperature \cite{Gol}. Because our
analysis focuses on the zero-Berry-curvature limit, how analogous
high-temperature scattering processes affect the LTET identified here
remains an important open question.

Although we have used TSCO as a representative VLC 2D material, the
underlying mechanism is more general. Our theory should apply broadly to VLC
systems in which $E_{\perp }$ controls exciton anisotropy. In materials with
nonzero exciton Berry curvature, the anisotropy-driven LTET studied here may
coexist and compete with Berry-curvature-induced transverse currents, as
well as with the layer counterflow response in twisted bilayers, due to the
long life time of excitons in VLC 2D materials \cite{SAYang,Ji}. Clarifying
how these channels evolve with $\theta $, $E_{\perp }$, and $T$, and whether
they can be cleanly disentangled experimentally, is an important direction
for future work.

In summary, using TSCO as a prototype VLC monolayer and twisted bilayer, we
show that $E_{\perp }$ directly controls exciton-dispersion anisotropy and
thereby induces unconventional LTET. In the monolayer, the response is
captured by the anisotropy-induced transverse conductivities $\sigma^{\mathrm{AS}}_{%
\mathrm{tra}}$ and $\alpha^{\mathrm{AS}}_{\mathrm{tra}}$, arising from the
antisymmetric part of diagonal elements in the conductivity tensor and strongly tunable by $%
E_{\perp }$. In the twisted bilayer, this anisotropy-driven response
coexists and competes with TREHCF, while the reduced symmetry of TSCO
additionally allows $\sigma^{\mathrm{S}}_{\mathrm{tra}}$ and $\alpha^{%
\mathrm{S}}_{\mathrm{tra}}$ from symmetric part of the conductivity tensor.
We further show that their interplay can be tuned efficiently by twisted
angle, electric field, temperature, and the direction of the in-plane
driving force, providing a practical route to isolate distinct transverse
exciton transport signals. These results establish $E_{\perp }$ as a
powerful knob for engineering exciton transport via VLC and suggest a
general strategy for controlling excitonic transport in a broad class of 2D
materials.

\begin{acknowledgments}
C. L. would like to thank W. Yao and B. B for useful discussions.
This work is supported by the National Natural Science Foundation of China (12504205),
the Fundamental Research Funds for Young Scholars of Jiangsu Province
(BK20250344), and the Fundamental Research Funds for the Central
Universities from China.
\end{acknowledgments}

\end{document}